\documentclass[conference]{IEEEtran}

\usepackage[utf8]{inputenc}   

\usepackage{graphicx}         
\graphicspath{ {images/} }

\usepackage{times}            
\usepackage{cite}
\usepackage{algpseudocode}
\usepackage{algorithm}
\usepackage[export]{adjustbox}
\usepackage[font=small,skip=10pt]{caption}
\usepackage{amsmath,amssymb,amsfonts}
\usepackage[T1]{fontenc}
\usepackage[section]{placeins}
\usepackage{amssymb}
\usepackage{textcomp}
\usepackage{xcolor}
\usepackage{comment}
\usepackage[english]{babel}
\usepackage[english]{algxpar}
\usepackage[section]{placeins}
\usepackage{makecell}

\ifCLASSINFOpdf
\else
\fi
\begin{document}
%

\title{Design and Analysis of Approximate Hardware Accelerators for VVC Intra Angular Prediction}


\author{\IEEEauthorblockN{Lucas M. Leipnitz de Fraga, Cláudio Machado Diniz}
\IEEEauthorblockA{PGMICRO, Institute of Informatics, Federal University of Rio Grande do Sul, Porto Alegre, Brazil \\
\{lucas.fraga,claudio.diniz\}@inf.ufrgs.br}
}


%


\maketitle

\begin{abstract}
The Versatile Video Coding (VVC) standard significantly improves compression efficiency over its predecessor, HEVC, but at the cost of substantially higher computational complexity, particularly in intra-frame prediction. This stage employs various directional modes, each requiring multiple multiplications between reference samples and constant coefficients. To optimize these operations at hardware accelerators, multiplierless constant multiplication (MCM) blocks offer a promising solution. However, VVC's interpolation filters have more than fifty distinct coefficients, making MCM implementations resource-intensive. This work proposes an approximation method to reduce the number of interpolation coefficients by averaging fixed subsets of them, therefore decreasing MCM block size and potentially lowering circuit area and power consumption. Six different MCM block architectures for angular intra prediction are introduced, in which five use the approximation method introduced in this work, and evaluate the trade-off between coefficient reduction and coding efficiency compared with a conventional multiplier architecture. Experimental results in ten videos demonstrate that only two MCM implementations exceed a 4\% BD-Rate increase and 2.6\% on average in the worst case, while two of the MCM implementations have circuit area reduction of 20\% and 44\%. For three of the architectures, parallel sample prediction modules were synthesized, showing a reduction of 30\% gate area compared to single sample processing units, and a reduction in energy consumption for two of the implementations.
\end{abstract}


\IEEEpeerreviewmaketitle

\section{Introduction}\label{introduction}

Video coding standards have become increasingly more complex to keep up with the demand for higher-quality real-time video applications. The Versatile Video Coding (VVC), developed by ISO/IEC and ITU-T \cite{VVC}, surpasses its predecessor H.265/HEVC in compression efficiency at the expense of higher computational complexity \cite{Bross2021}. A detailed analysis of computational complexity comparing VVC and HEVC demonstrated that VVC ensures a 44\% bitrate savings compared to its predecessor while maintaining the same visual quality, although with a runtime up to ten times higher \cite{Siqueira2020}.

One of the most expensive stages of a video encoder is the intra-frame prediction \cite{intra_in_vvc}, which seeks to find redundancies in a single frame. It has 93 directional modes, which are all tested to find the best compression result. These modes are composed of interpolation filters, with multiple multiplications of reference samples (variables) by coefficients (constants). It is essential to optimize the multiplication process at hardware level, since it will occur several times during the prediction.

Field Programmable Gate Arrays (FPGAs) have specific units for multiplication in their programmable fabric. Hence, it is not useful to simplify multiplications since the multipliers are already fabricated \cite{fpga_arq}. However, in Application Specific Integrated Circuits (ASICs), optimization in multiplications leads to lower area and power dissipation. One way of implementing multiplication by constants is using multiplierless multiple constant multiplication (MCM) blocks, composed only of addition and shift operations.

However, VVC has 57 distinct coefficients, which can be burdensome to process all multiplications, even for MCM blocks. Our work proposes to approximate the prediction equation by reducing the number of interpolation coefficients, aiming to reduce the MCM block size, possibly contributing to the reduction of circuit area and power consumption. Nevertheless, reducing the number of coefficients may have negative impacts on the compression efficiency. Therefore, a deeper analysis on compression efficiency drop is necessary.

Works in literature have already proposed the use of MCM blocks in angular intra prediction or multiplications as adders and shifters with distinct approximation methods. In \cite{hevc_arq}, a reference sample approximation method was introduced to facilitate the reuse of prediction results for HEVC. The architecture uses an MCM datapath, with BD-rate average results of just 0.46\%, with an architecture 10 times smaller than other works in the literature. In \cite{asic_arq}, a heuristic of most frequently chosen modes is presented, managing to reach the smallest area between two other works in the literature, with an increase of 8.62\% average BD-rate.

This work introduces a new approximation method for the VVC intra angular prediction for the \textit{luma} component, by replacing the interpolation coefficients by an average of a fixed number of coefficients, to reduce the complexity of MCM block. Six different MCM blocks architectures were designed for VVC intra angular equations, with variations for calculating multiple samples in parallel for 4 of them. To the best of our knowledge, no other work has proposed a similar approximation strategy. 

Results of experiments with ten video sequences and the five implementations of different approximation methods show that only two results remained above 4\% of the BD-Rate, with the average varying between 0.3\% and 2.6\%. Two of the MCM implementations have a lower gate area, i.e. 20\% and 44\% area reduction compared to the conventional multiplier approach. For three of the architectures, experiments with parallel sample prediction modules were conducted, and show a reduction of 30\% gate area compared to single sample processing units.

\section{Background}

\subsection{Intra-Frame Prediction in VVC}\label{intra_background}

This section briefly discusses intra-frame prediction in VVC. More details on intra prediction, please refer to \cite{VVC}.

Initially, an intra frame is divided into multiple Coding Tree Units (CTUs), which are further divided in rectangular Coding Units (CUs). Each CU consists of a single Transform Block (TB), except in the cases of Intra Subpartition (ISP) or implicit splitting, where intra prediction takes place \cite{intra_in_vvc}.

To perform intra prediction, the encoder relies on reference samples, which may change from mode to mode, but in all cases, are a subset of a row and column of previously predicted neighboring blocks, located, respectively, above and on the left of the current prediction block.

Intra directional modes are composed of intra angular (modes 2 to 66) and Wide Angular Intra Prediction (modes -1 to -14 and 66 to 80), or WAIP. To select the optimal intra mode for each CU, the encoder finds the minimal rate distortion cost (RD-cost), which then encodes the unit with the chosen mode.

Directional modes for the luminance component, the focus of this work, are calculated with 4-tap filters, consisting of a sum of multiplications of four samples by different constant coefficients. These are chosen from two tables of 32-row by 4-column, each representing two different filters, and are chosen based on the mode and predicted reference position. Equation \ref{eq:angular} is an example of an angular prediction. 

\begin{equation}
p(x)(y) = Clip(((\sum_{i=0}^{3}f[k][i]*r[ x + i_0 + i ]) + 32) >> 6)
\label{eq:angular}
\end{equation}

In (\ref{eq:angular}), \textit{p} is the predicted sample, with \textit{x} and \textit{y} being its position relative to the TB, and \textit{r} the reference vector. Both $k \in \{0, 1, ..., 31\}$ and $i_0$ depend on \textit{y} and the angular mode. They represent, respectively, the interpolation filter coefficients table line and the position of the first reference sample in the reference vector. The filter coefficients table \textit{f} are signed integers that either represent a DCT-based interpolation filter, referred to as \textit{fC}, or a 4-tap smoothing interpolation filter, as \textit{fG} \cite{intra_in_vvc}.

\subsection{Multiple Constant Multiplication}\label{mcm_background}
 
Multipliers are considerably costly in hardware, and many multiplications are used to compute intra-frame prediction filters. To avoid them, the filter multiplications can be implemented as a combination of shifts, additions and negations.
 
Multiplierless Multiple Constant Multiplication (MCM) is a problem that involves finding the optimal way to calculate a set of products between the same variable and multiple fixed-point constants, using only additions, negations, and shifts. In this work, the \textit{Hcub} heuristic \cite{mcm} and the \textit{Spiral} software \cite{spiral} were used to produce the MCM blocks that calculate the interpolation filters.

\section{Proposed Interpolation Filters Approximation}\label{aproximation}

Aiming to reduce the size of MCM blocks, in this section, we present our strategy to reduce the number of coefficients required to calculate the intra angular prediction. It is an approximation for the RD-cost calculation, which means that it is only applicable for deciding the best prediction mode, and not for generating the final bitstream, which uses precise intra prediction. This is done because approximating intra prediction used to generate the final bitstream results in a mismatch of the video encoder and decoder, since the video decoder must be precise to be compatible with the VVC standard.

The strategy is based on the average sequences of the \textit{n} coefficients from the columns of the filter coefficient table. The higher the value of \textit{n}, the fewer different coefficients the table will have, with a higher impact on RD-cost calculation.

\begin{figure}[H]
    \centering
    \includegraphics[width=0.75\linewidth]{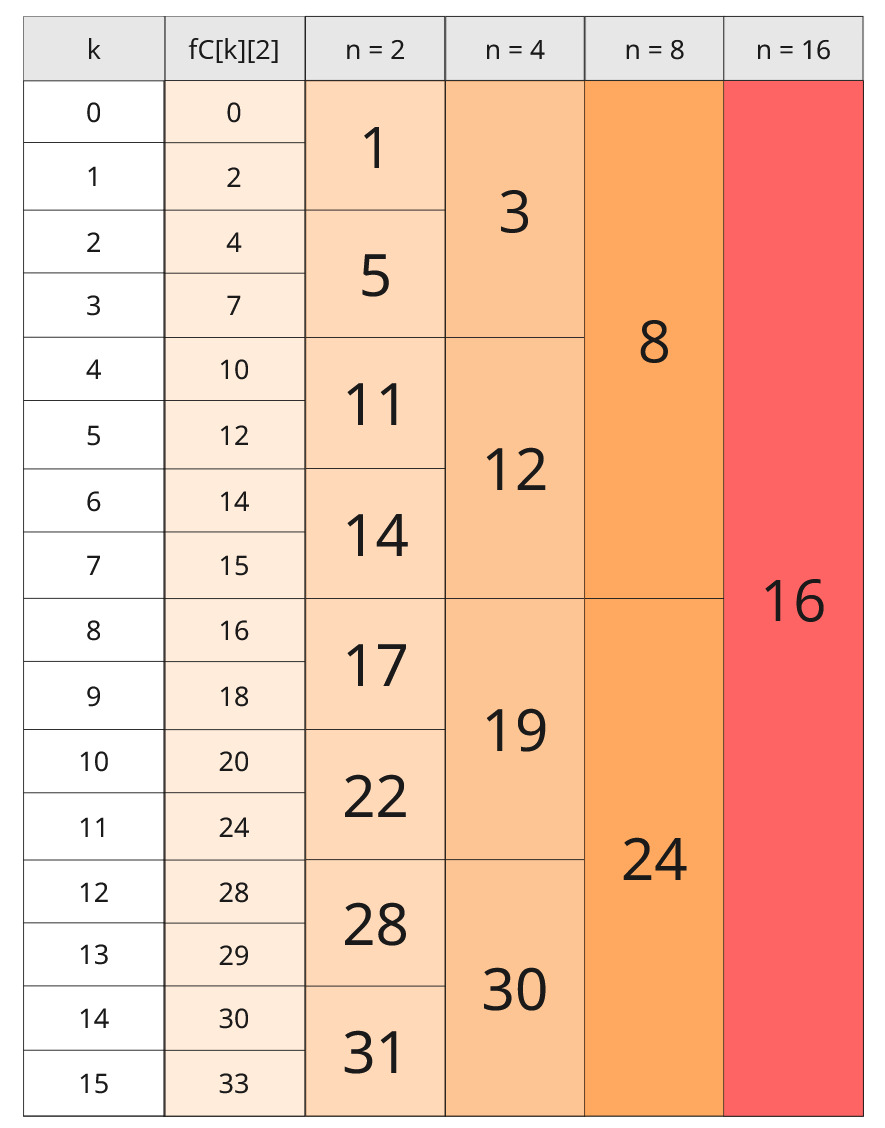}
    \caption{Example of interpolation filters approximation}
    \label{fig:aprox_table}
\end{figure}

Fig. \ref{fig:aprox_table} exemplifies the strategy with the first 16 lines of the third column of the \textit{fC} filter table, for four different values of \textit{n}. When \textit{n} equals 2, the first and second values of the column are represented by the average of the first and second original values, as well as the third and fourth by the average of the third and fourth original values, and so on. The same rule applies when \textit{n} equals 4, 8, 16 and 32, but averaging sequences of four, eight, sixteen and thirty-two values instead of two. For the rest of the columns, the same rule applies.

\section{Proposed Architectures}\label{arch}

\begin{figure}
    \centering
    \includegraphics[width=1\linewidth]{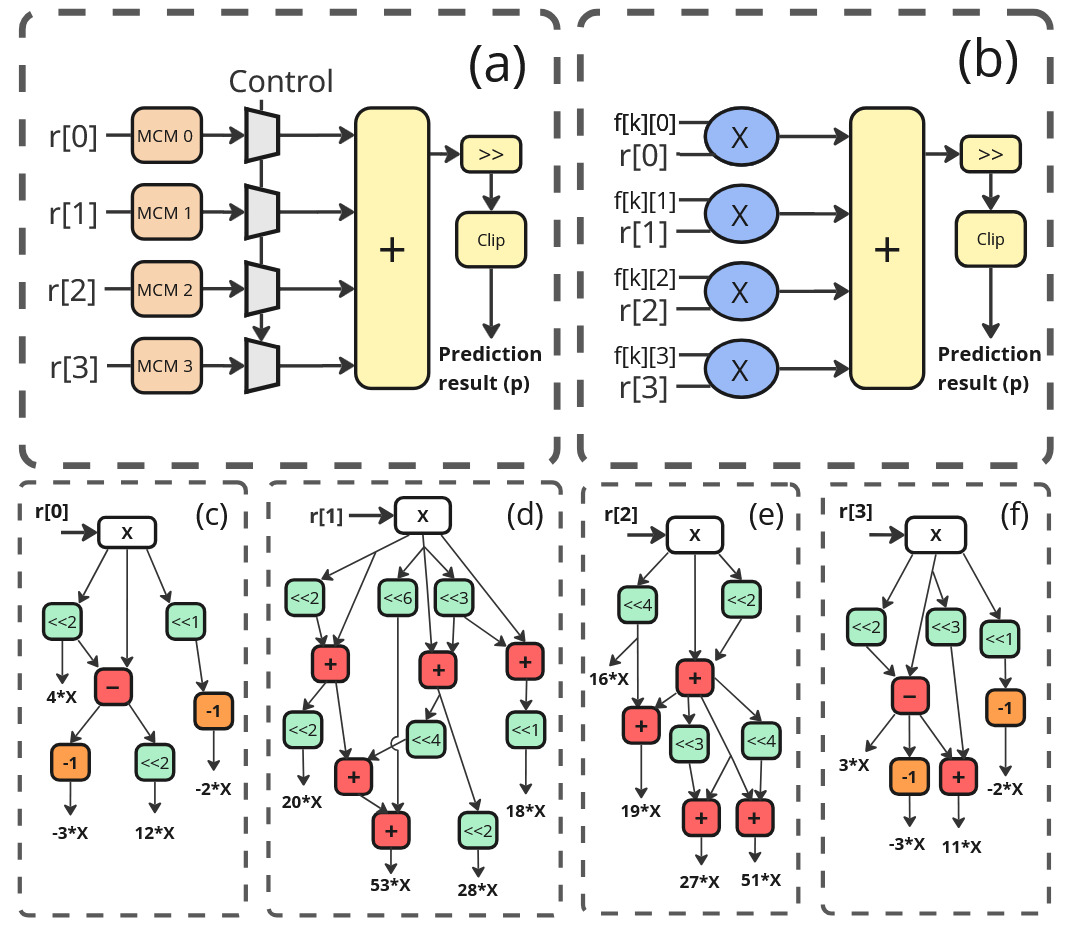}
    \caption{Proposed architectures}
    \label{fig:arch}
\end{figure}

In this section, we present two different types of architectures for calculating the prediction for all angular modes and block sizes, as well as all seven implementations of these architectures. Fig. \ref{fig:arch} shows the two types of architectures, while all implementations are summarized in Table \ref{tbl:arch}.

Fig. \ref{fig:arch}(a) shows the first architecture type, which uses MCM blocks to calculate multiplications by constants, in addition to multiplexers and control bits to choose which of them will be sent to sum, shift and clip steps. Six variants of this architecture were implemented, in which five use the approximations presented in Section \ref{aproximation} to select a smaller number of coefficients for the MCM blocks and the latter employs all the coefficients for precise calculation. Fig. \ref{fig:arch}(c) to \ref{fig:arch}(f) show, as an example, the four MCM blocks, from 0 to 3 respectively, for \textit{mcm\_n16}, changing according to the architecture variation.

Each block calculates all multiplications of the reference with the selected coefficients from the respective column of the two filters. For example, block \textit{MCM 2} calculates the multiplications of \textit{r[2]} by constants 16, 51, 19 and 27, where the first two are from filter \textit{fC} and the last two are from filter \textit{fG}. Repeated values are calculated only once. Multiplications by zeros and ones are not calculated in MCM blocks, and the references are sent directly to the adder for the latter case.

The last implementation is based on the second architecture, represented by Fig. \ref{fig:arch}(b), which uses conventional multipliers that receive, as input, the values of the references and coefficients and send them to subsequent steps. The coefficients are read from a read-only memory (ROM), which has stored all the VVC filter coefficients. This implementation also has precise prediction.

\begin{table}[h]
    \caption{Summary of the architecture implementations.}
    \centering
    \resizebox{1.0\columnwidth}{!}{%
        \begin{tabular}{|p{1.8cm}|p{1.5cm}|p{1.5cm}|p{2cm}|p{2cm}|}
          \hline
          \thead{Implementation} & \thead{Architecture} & \thead{Average (\textit{n})} & \thead{Multiplication\\ Method} & \thead{Predicted\\ Sample}\\
          \hline
          \thead{mcm\_n32} & \thead{\textit{(a)}} & \thead{32} & \thead{MCM Blocks} &
          \thead{Approximate} \\
          \hline
          \thead{mcm\_n16} & \thead{\textit{(a)}} & \thead{16} & \thead{MCM Blocks} &
          \thead{Approximate} \\
          \hline
          \thead{mcm\_n8} & \thead{\textit{(a)}} & \thead{8} & \thead{MCM Blocks} &
          \thead{Approximate} \\
          \hline
          \thead{mcm\_n4} & \thead{\textit{(a)}} & \thead{4} & \thead{MCM Blocks} &
          \thead{Approximate} \\
          \hline
          \thead{mcm\_n2} & \thead{\textit{(a)}} & \thead{2} & \thead{MCM Blocks} &
          \thead{Approximate} \\
          \hline
          \thead{mcm\_precise} & \thead{\textit{(a)}} & \thead{-} & \thead{MCM Blocks} &
          \thead{Precise} \\
          \hline
          \thead{mult} & \thead{\textit{(b)}} & \thead{-} & \thead{Multipliers} &
          \thead{Precise} \\
          \hline
        \end{tabular}}
    \label{tbl:arch}
\end{table}

\begin{figure}
    \centering
    \includegraphics[width=0.75\linewidth]{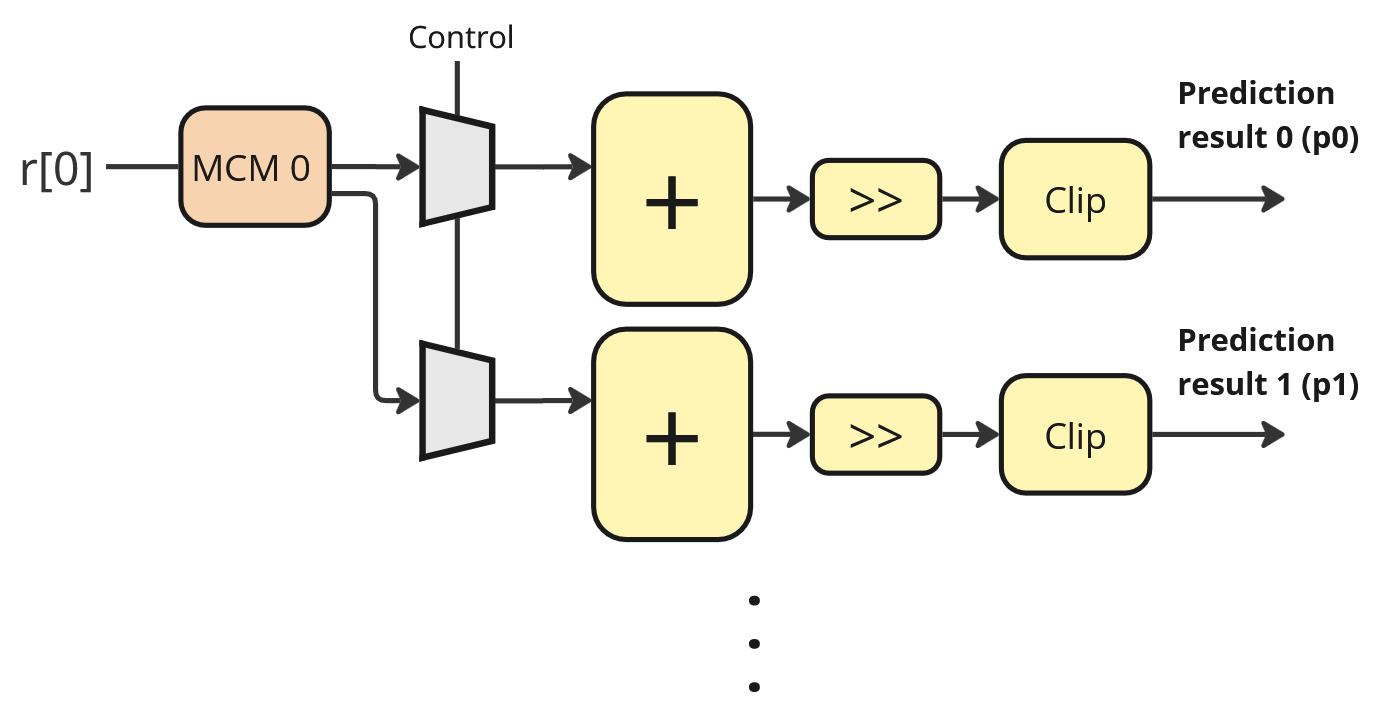}
    \caption{MCM architecture for parallel prediction}
    \label{fig:mcm}
\end{figure}

All implementations were described in VHDL/Verilog, with the MCM blocks automatically generated by \textit{Spiral} \cite{spiral} in Verilog language. For \textit{mcm\_n32}, \textit{mcm\_n16}, \textit{mcm\_n8} and \textit{mult} implementations, variations for the prediction of 4, 8, 16, 32 and 64 parallel samples were also implemented. The goal is to use the MCM blocks to process various multiplications by the same reference sample, using them in multiple predictions at the same time, and compare with the multiplier solution. Fig. \ref{fig:mcm} exemplifies this case for a sample \textit{r[0]} in the calculation of two parallel samples. The advantage is that \textit{MCM 0} is less complex than its two counterparts concerning two parallel prediction modules of only one sample, as shown in Fig. \ref{fig:arch}(a), since it can have more shared adders and shifters for all the multiplications.

\section{Results}\label{results}

In this section, we present the evaluation on compression efficiency of the approximate implementations related to the precise implementations (\textit{mcm\_precise} or \textit{mult}), using the Bjøntegaard Delta Rate (BD-Rate) metric. Then, we present the logic synthesis results of all seven implementations and a comparison of area and power.

To calculate BD-Rate, ten randomly chosen videos from VVC common test conditions were encoded, with quantization parameters of 22, 27, 32, and 37 \cite{common_test_conditions}. The videos were encoded in the original VVC Test Model (VTM), the reference software for VVC, and a modified version that implements each one of the five approximations presented in Section \ref{aproximation}. Each video was encoded with its original resolution and with one second of duration calculated by the recommended frame rate of each video sequence (\textit{e.g.} 30 frames for a 30 fps rate).

\begin{figure}[h]
    \centering
    \includegraphics[width=1\linewidth]{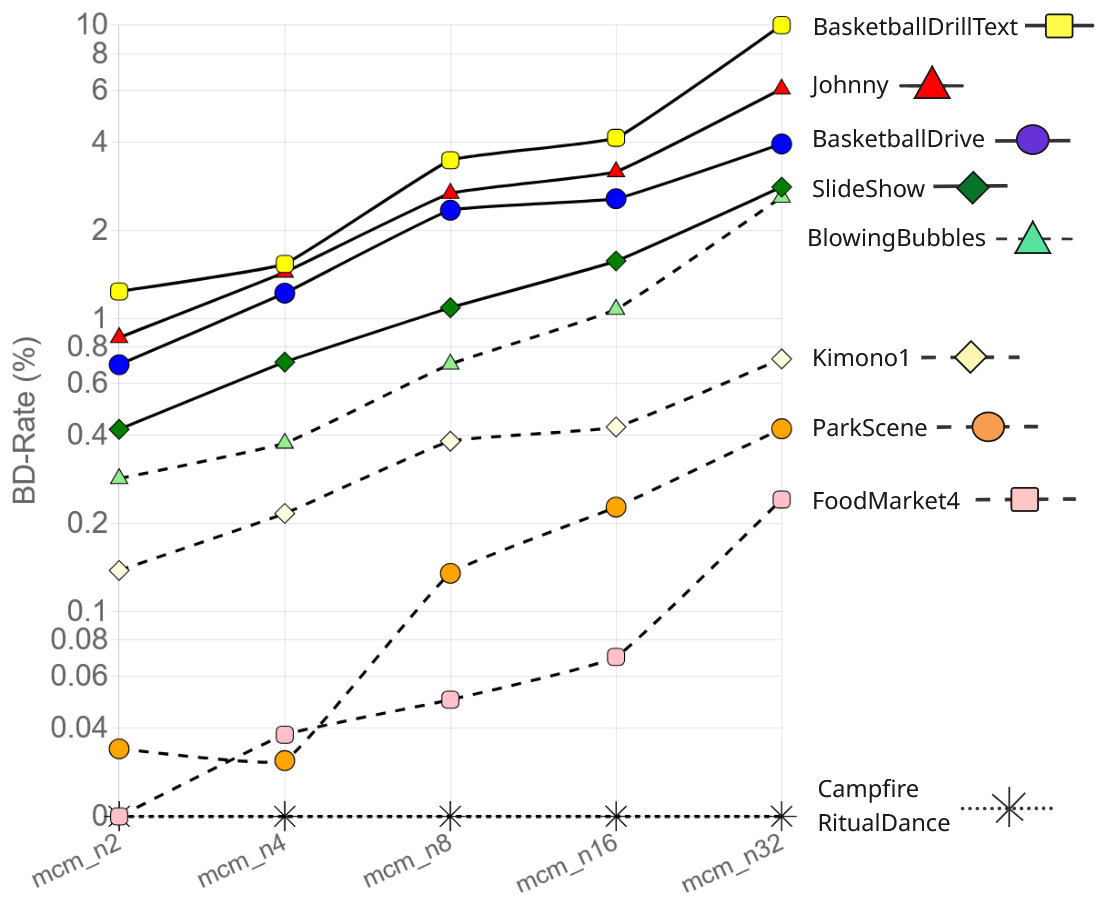}
    \caption{BD-Rate for each video on each approximate implementation}
    \label{fig:bd-rate2}
\end{figure}

Fig. \ref{fig:bd-rate2} shows, on a logarithmic scale, the BD-Rate increase results for each video for each approximation. Campfire and RitualDance videos did not have relevant variations in BD-rate, indicating no significant compression efficiency drop of the approximate implementations related to precise implementation. However, the other eight videos had a higher percentage as the \textit{n} value increases. However, the majority were still able to stay below the 5\% mark, showing promising approximation results. This is expected, since, the higher the value of \textit{n}, the fewer distinct coefficients for calculating the interpolation, resulting in suboptimal choices of prediction modes. Furthermore, of the five worst-performing videos, four have a resolution lower than Full HD, indicating that lower resolution videos are the most affected by the approximations.

The logic synthesis was done using the Cadence Genus tool for the ST 65 nm technology. To avoid high operating frequencies and power costs, the synthesis was performed considering a \textit{hardware} for the prediction of 512 parallel samples, composed of multiple units in parallel based on the architecture implementations presented. 

VVC has 5 quadratic and 12 rectangular CU sizes \cite{asic_arq}. For each quadratic CU in a frame region of $64 \times 64$ samples, ($64\times64)/(512)\times 65$ (number of angular modes) cycles are needed, multiplied by the 5 possible quadratic blocks dimensions. The same calculation is made for the 12 possible rectangular blocks, but considering the 28 WAIP modes. Since there are 506.25 CUs in a 1080p frame ($(1920\times1080)/(64\times64)$), the architecture requires 2,677,050 cycles to process all CU sizes in a frame. Then, to process 30 frames per second, an operating frequency of 80 MHz is needed.

Table \ref{tbl:sint} compares the unit and total (unit times 512) area and power for all the different hardware accelerators, and BD-Rate for the approximate ones. The area is represented in gate count equivalent to the 2-input NAND gate, which is the total area divided by the area of a 2-input NAND gate synthesized in the same technology.  Each hardware is composed of 512 equal units of its implementation presented in Section \ref{arch}. 

\begin{table}
    \caption{Logic synthesis results and average BD-rate for each accelerator}
    \centering
    \resizebox{1.0\columnwidth}{!}{%
        \begin{tabular}{|p{1.8cm}|p{2cm}|p{2cm}|p{1.5cm}|p{
        1.7cm}|p{1.5cm}|}
          \hline
          \thead{Hardware} & \thead{Total Area \\(gate-equivalent)} & \thead{Unit Area \\(gate-equivalent)} & \thead{Total Power \\($m W$)} & \thead{Unit Power \\($\mu W$)} & \thead{BD-Rate \\Average (\%)}\\
          \hline
          \thead{mcm\_n32} & \thead{148480} & \thead{290} & \thead{257.024} & \thead{502} & \thead{2.678}\\
          \hline
          \thead{mcm\_n16} & \thead{266240} & \thead{520} & \thead{401.920} & \thead{785} & \thead{1.332}\\
          \hline
          \thead{mcm\_n8} & \thead{355840} & \thead{695} & \thead{488.960} & \thead{955} & \thead{1.074}\\
          \hline
          \thead{mcm\_n4} & \thead{497664} & \thead{972} & \thead{587.264} & \thead{1147} & \thead{0.556} \\
          \hline
          \thead{mcm\_n2} & \thead{794112} & \thead{1551} & \thead{785.920} & \thead{1535} & \thead{0.373} \\
          \hline
          \thead{mcm\_precise} & \thead{894976} & \thead{1748} & \thead{818.176} & \thead{1598} & \thead{-} \\
          \hline
          \thead{mult} & \thead{332800} & \thead{650} & \thead{384} & \thead{750} & \thead{-} \\
          \hline
        \end{tabular}}
    \label{tbl:sint}
\end{table}

Considering all implementations, only \textit{mcm\_n16} and \textit{mcm\_n32} were able to reduce around 20\% and 44\%, respectively, the total area, but the first one is still worse in total power. Accuracy is another advantage of \textit{mult}, nonetheless, both \textit{mcm\_n16} and \textit{mcm\_n32} had small increases of only one and two points in the BD-Rate average of all videos tested. 

As mentioned in Section \ref{arch}, hardware accelerator units for parallel samples computing were synthesized following the previous rule of predicting 512 samples in one cycle. For example, 8 units that compute 64 parallel samples are needed to compute all samples. The area comparison is presented in Table \ref{tbl:sint_2}, while Table \ref{tbl:sint_3} presents the power. 

Firstly, there is a reduction of up to 30\% in area, on average, the greater the number of samples predicted in parallel, for all four implementations. Secondly, the implementation \textit{mcm\_n8} had a final area of 1784 logic gates smaller compared to \textit{mult}, which was originally 23040 gates larger. Finally, the power decreased for the \textit{mcm\_n8} and \textit{mcm\_n16} cases, but increased for the \textit{mcm\_n32} and \textit{mult} cases, and remained approximately constant as the samples calculated in parallel increased.

\begin{table}[H]
    \caption{Gate count for prediction unit with different quantities of samples in parallel}
    \centering
    \resizebox{1.0\columnwidth}{!}{%
        \begin{tabular}{|p{2cm}|p{1.5cm}|p{1.5cm}|p{1.5cm}|p{1.5cm}|p{1.5cm}|}
          \hline
          \thead{Hardware} & \thead{4 Samples} & \thead{8 Samples} & \thead{16 Samples} & \thead{32 Samples} & \thead{64 Samples}\\
          \hline
          \thead{mcm\_n32} & \thead{125396} & \thead{117856} & \thead{113984} & \thead{112068} & \thead{111516}\\
          \hline
          \thead{mcm\_n16} & \thead{198612} & \thead{186296} & \thead{181280} & \thead{176508} & \thead{175164}\\
          \hline
          \thead{mcm\_n8} & \thead{304800} & \thead{289116} & \thead{279876} & \thead{276088} & \thead{276032}\\
          \hline
          \thead{mult} & \thead{288000} & \thead{284192} & \thead{280512} & \thead{278960} & \thead{277816}\\
          \hline
        \end{tabular}}
    \label{tbl:sint_2}
\end{table}

\begin{table}[H]
    \caption{Total power ($mW$) for prediction with different quantities of samples in parallel}
    \centering
    \resizebox{1.0\columnwidth}{!}{%
        \begin{tabular}{|p{2cm}|p{1.5cm}|p{1.5cm}|p{1.5cm}|p{1.5cm}|p{1.5cm}|}
          \hline
          \thead{Hardware} & \thead{4 Samples} & \thead{8 Samples} & \thead{16 Samples} & \thead{32 Samples} & \thead{64 Samples}\\
          \hline
          \thead{mcm\_n32} & \thead{284} & \thead{268} & \thead{232} & \thead{260} & \thead{262}\\
          \hline
          \thead{mcm\_n16} & \thead{332} & \thead{328} & \thead{332} & \thead{320} & \thead{332}\\
          \hline
          \thead{mcm\_n8} & \thead{448} & \thead{432} & \thead{408} & \thead{420} & \thead{424}\\
          \hline
          \thead{mult} & \thead{400} & \thead{384} & \thead{348} & \thead{388} & \thead{412}\\
          \hline
        \end{tabular}}
    \label{tbl:sint_3}
\end{table}

\section{Conclusions}

The approach introduced in this work is intended to reduce the complexity of MCM blocks in the angular prediction of VVC. The presented synthesis shows area reduction of around 20\% and 44\% for the cases of averaging 16 and 32 coefficients, compared with a conventional multiplier architecture, with a compression efficiency drop of 2.6\% average (an increase in BD-Rate) in the worst case.
For the parallel processing units, there is a reduction of 30\% compared to single processing units, with the averaging of 8 values reducing its area compared to the multiplication architecture in 1784 gates, and a reduction in energy consumption for the 16 and 8 averages.


\section*{Acknowledgment}

This work is supported in part by Coordenação de Aperfeiçoamento de Pessoal de Nível Superior - Brazil (CAPES) - Financial Code 001, CNPq, and FAPERGS. We also thank the Brazilian Microelectronics Society (SBMicro) for providing access to commercial ASIC tools through the APCI program (Programa de Apoio a Projeto de Circuitos Integrados em Universidades).



%

\bibliographystyle{IEEEtran}
\bibliography{references}

\end{document}